\documentclass[twocolumn,aps,preprintnumbers,amsmath,amssymb,footinbib,prl]{revtex4}

\usepackage{graphicx}
\usepackage{dcolumn}
\usepackage{bm}
\usepackage{epsfig}

\usepackage{amsmath}
\usepackage{amssymb}
\usepackage{color}

\def\fun#1#2{\lower3.6pt\vbox{\baselineskip0pt\lineskip.9pt
  \ialign{$\mathsurround=0pt#1\hfil##\hfil$\crcr#2\crcr\sim\crcr}}}

\input epsf

\newcommand{\be}{\begin{equation}}
\newcommand{\ee}{\end{equation}}
\newcommand{\bea}{\begin{eqnarray}}
\newcommand{\eea}{\end{eqnarray}}

\begin{document}


\title{Split-UED and Dark Matter}

\author{Seong Chan Park, Jing Shu}
\affiliation{\small Institute for the Physics and Mathematics of the Universe, The University of Tokyo, Chiba $277-8568$, Japan
}

\begin{abstract}
Motivated by the recent observation of the high energy electron and positron excesses in cosmic ray by PAMELA and ATIC/PPB-BETS, we suggest an anomaly-free scenario for the universal extra dimension that localizes the SM quarks and splits the spectrum of KK quarks from KK leptons. When the SM quarks are ``well localized" at the boundaries, the most stringent bound of the model ($1/R > 510$ GeV) comes from the resonance search for the Tevatron dijet channels. Even at the early stage of LHC, one can discover the second KK gluon for masses up to 4 TeV.   
\end{abstract}

\pacs{draft}

\maketitle

\preprint{IPMU 09-0003}

\noindent {\em Introduction--} One of the best motivated ways of extending the standard model (SM) is to embed the theory in higher dimensions. The ``direct proof" of dark matter (DM) \cite{Clowe:2006eq} certainly brings our attention to a particular form of extra dimensional model \cite{Antoniadis:1990ew}, the universal extra dimension (UED) \cite{Appelquist:2000nn},  in which the lightest Kaluza-Klein particle (LKP) arises as a natural candidate for dark matter \cite{Servant:2002aq} thanks to the Kaluza-Klein (KK)  parity (for the comprehensive recent review, see e.g. \cite{Hooper:2007qk}). Indeed the ATIC collaboration recently claimed that their observation of the excess and the sharp drop in the cosmic ($e^++e^-$) spectrum around 300-800 GeV can be naturally understood by the electrons and positrons from the LKP annihilation assuming a large ``boost factor" in the galactic halo \cite{ATIC}. The result is essentially consistent with the PPB-BETS result \cite{PPB-BETS}. It is even more interesting to notice that the same source of positrons can also explain another interesting observation by PAMELA \cite{Cholis:2008wq}, where the ratio of the positron flux over the sum of the electron and positron fluxes in the energy range 10-85 GeV is much higher than the standard astrophysical expectation \cite{PAMELA}. However, one should note that no excess in antiproton flux has been observed by the same experiment and thus the UED needs to be modified \footnote{In Ref. \cite{Hooper:2009fj} the authors claimed that the absence of an excess in anti proton signal can be consistent with  MUED by choosing a particular galactic propagation model. However it is still desirable to have a mechanism to reduce the hadronic branching fraction.}.  Because of the characteristic degeneracy in the KK spectrum, UED predicts that a pair of the LKP, the first KK photon, annihilates not only to leptons but also to quarks at a comparable rate. The main purpose of this letter is to provide a simple way of modifying UED in such a way that the KK dark matter annihilates mainly into leptons and the hadronic production is suppressed by the heavier KK quarks as $\langle \sigma v \rangle_{q\bar{q}} \propto m_{\gamma^1}^2/(m_{\gamma^1}^2+m_{q^1}^2)^2$.

Conventional wisdom is that KK parity is available only in the case when all the fields are propagating through the bulk. However, we notice that KK parity remains a good symmetry even when some fields are (quasi-) localized at the boundaries if their profiles respect the inversion symmetry about the midpoint. As it is clearly seen in Fig. 1, the inversion symmetry about the midpoint ($y=0$), and thus the KK parity, is respected even in the case when the quarks are quasi-localized at the boundaries ($y=\pm L$).
As we will see in detail below, the quasi-localization makes the KK quarks heavier, and consequently their contribution to the dark matter annihilation becomes more suppressed. 
We suggest call this set-up split-UED, as the spectrum of the KK quarks is quite split from the others, and the profile of each quark in the fifth dimension is also quite split towards the two boundaries \footnote{In terms of particle spectrum, split-UED is quite similar to split-SUSY \cite{ArkaniHamed:2004fb} and these two models share many features in the collider phenomenology.} \footnote{Adding a brane kinetic term that is symmetric on the boundary branes will also split mass spectrum and preserve the KK parity, and this is considered in Ref. \cite{Flacke:2008ne} for gauge boson and Higgs.}.

This letter is organized as follows. First, we provide a concrete field theoretic method of quasi-localizing fermions on the boundaries while keeping the KK parity intact.  
The realistic model of KK dark matter is constructed by embedding the SM in the set-up where the quarks are quasi-localized at the boundaries. We show that our setup is anomaly free. 
After considering the current bounds from electroweak data and flavor physics, we discuss the LHC  phenomenology, which is quite distinct from the conventional minimal UED (MUED). 

\begin{figure}
\includegraphics[scale=0.30, angle=0 
]{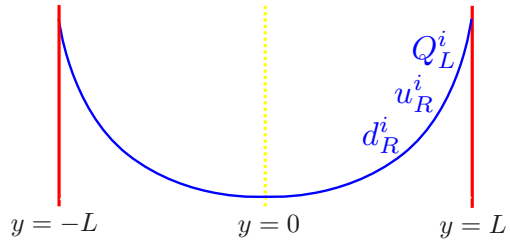}  
\caption{\label{fig1}The profile of wave functions of the quasi-localized quarks in the extra dimension.
The profile clearly respect the inversion invariance about the midpoint ($y=0$), and is localized toward
the end points ($y=\pm L$).}
\end{figure}

\noindent {\em   A Field Theory Realization--}
Here we present an explicit field theory mechanism to localize fermion zero modes at the boundaries (or fixed points) in higher dimensions in such a way that the KK parity is conserved. We start our setup by considering 5D fermions on an orbifold along the fifth dimension with the boundary points $y = \pm L(=\pm \pi R/2) $. The 5D bulk Lagrangian is given by the form 
\begin{eqnarray}
\label{eq:Lag}
S= \int d^5 x \Big[\frac{i}{2} (\bar{\Psi}_I \Gamma^M\overleftrightarrow{\partial_M} \Psi_I) - \lambda_{IJ} \Phi(y) \bar{\Psi}_I \Psi_J \Big],
\end{eqnarray}
where $\overleftrightarrow{\partial_5} = \overrightarrow{\partial_5}-\overleftarrow{\partial_5}$, with the arrows indicating the direction of action of the differential operator, $M=0,1,2,3,5$ and $\Gamma^M = (\gamma^\mu, i \gamma_5)$, which satisfies the Clifford algebra in 5D: $\{\Gamma^M, \Gamma^N \}=2\eta^{M N}$, $\eta^{M N} = diag(1,-1,-1,-1,-1)$. $\lambda_{IJ}$ is the Yukawa coupling between the background scalar field $\Phi(y) $, fermion $\Psi_I(y)$ and $\Psi_J(y)$ and $I, J$ are the flavor indexes
\footnote{We will assume the Yukawa interactions diagonal 
$\lambda_{IJ} = \lambda \delta_{IJ}$ to avoid the potential flavor problems.}. 
Here if we choose an anti-symmetric background profile $ \Phi(-y) = - \Phi(y)$, we can see that the KK parity, which is the inversion symmetry about the midpoint $y=0$, is a good symmetry of the Lagrangian under which the fermion field transforms as $\Psi(x, -y) \rightarrow \pm \gamma_5 \Psi(x, y) $. The simplest bulk mass profile we can consider is the a step function $\lambda \langle \Phi(y)  \rangle = m(y) = \mu \epsilon(y)$, where the $\epsilon(y)$ is defined to be $+1$ for $0<y<L$ and $-1$ for $-L<y<0$. The origin of such a bulk mass profile could be understood as a particular limit of a double kink profile in a similar way to the orbifold setup in Ref. \cite{kink, Grossman:1999ra}. In this case, the allowed term $\lambda_H \Phi^2 H^\dag H$ can only contribute to the Higgs mass by $\lambda_H \langle \Phi \rangle^2$ in a $y$-independent way.

A bulk fermion is now KK expanded by basis eigenmodes which are sub-divided into even(symmetric) and odd(antisymmetric) modes under KK parity.  
A convenient way of expressing the KK decomposition for $Z_2$ even/odd ($\Psi_\pm$) mode is in the form \cite{Agashe:2007jb}:
\bea
\label{KKdecomp}
\Psi_+(x,y) &=& \sum_{n^+, n^-} g_{n^+}(|y|) \chi_{n^+}(x) + \epsilon(y) g_{n^-}(|y|) \chi_{n^-}(x), \nonumber \\
\Psi_-(x,y) &=& \sum_{n^+, n^-} \epsilon(y) f_{n^+}(|y|) \psi_{n^+}(x)+  f_{n^-}(|y|) \psi_{n^-}(x), \nonumber \\
\eea
where the label $n^\pm$ here stands for the $n$th KK modes with the even/odd KK parity.  
The profiles satisfy the following coupled, first-order equations of motion in $(0,L)$:  $\partial_y g_n + \mu g_n - m_n f_n= 0$ and $\partial_y f_n - \mu f_n + m_n g_n= 0$
with each profile $g_{n^+}$, $g_{n^-}$, $f_{n^+}$, $f_{n^-}$ satisfying the $(+,+)$, $(-,+)$, $(-,-)$, $(+,-)$ boundary condition at $y = (0,L)$, respectively. Once the solution for $y \subset [0,L]$ is obtained, the solution to the whole space is determined from Eq. (\ref{KKdecomp}) thanks to the symmetry. The solution for the zero mode in the interval $[0, L]$ is $g_{0^+}(y) = N_+ \exp (\int_0^y \lambda  \langle \Phi(s) \rangle ds) = N_+ \exp(\mu y)$. For the massive modes, when $\mu>0$, the profiles $g_{n^+}$ and $f_{n^-}$ were a combination of cosine and sine functions, while $f_{n^+}$ and $g_{n^-}$ were sine functions because of the $(-)$ boundary condition at $y=0$. The mass of the nth KK mode is given by $m_{n} = \sqrt{\mu^2 + k_{n}^2}$, where $k_{n^+} = {n \pi }/{L}$ for the KK even modes and $k_{n^-}$ is the nth solution of the equation $k_{n^-} = -\mu \tan(k_{n^-} L) $ for the KK odd modes, so that $k_{n^-}$ increases from $(n-1/2) \pi / L$ to $n \pi / L$ when $\mu$ increases from 0 to $+ \infty$. In this case, in the limit of $\mu \rightarrow + \infty$, all KK modes could be decoupled, and the zero mode is completely localized at the boundary $y = \pm L$ \footnote{In case of $\mu<-1/L$, the first KK mode is a special solution whose mass can be  approximated as $m_1 \approx 2 |\mu| e^{\mu L}$, which tends to zero in the $\mu \rightarrow - \infty$ limit.}.

\noindent {\em Embeding the Standard Model and the 5D Anomaly Cancellation--} 
The above setup could be used to quasi-localize any fermion zero modes along the fifth dimension. The lepton sector is untouched by putting $\lambda_{\rm lepton}=0$ to evade the stringent bounds from LEP and other low energy experiments. We choose $\lambda_{\rm quark}$ universally for all quarks and the massive KK quarks could be decoupled from the theory, so the dark matter candidate $\gamma_1$ will dominantly pair annihilate into lepton pairs. To avoid the very light 1st KK quark, we choose $\mu > + 1/L$ or $< - 1/L$ if we embed the SM quarks in the $\Psi_+$ or $\Psi_-$ component. In either case, one can see that the zero mode is quasi-localized at the boundary $y = \pm L$.

One immediate consequence of localizing the SM quarks is the violation of KK number conservation in the quark sector, which gives a tree level coupling of the KK even field to the SM quarks. For the KK gauge bosons, the effective coupling between the nth $(n>0)$ KK even gauge boson and the SM quarks could be obtained by integrating out the fifth dimension:
$g_{q^0-q^0-A^{2n}} =  \sqrt{2}g_0 {\cal F}_{2n}(\mu L)$ where $g_0$ is the SM gauge interaction between $q^0$ and $A^{0}$, and the dimensionless function ${\cal F}(\mu L)$ is given by:
\begin{eqnarray}
{\cal F}_{2n}(x)\equiv \frac{2 x}{1-e^{2 x}}\int_0^1 ds e^{2 x s} \cos (\pi n s),
\end{eqnarray}
where the dimensionless variable $s=y/L$, is introduced. Note that the function approaches $(-1)^{n+1}$ when $x=\mu L \gg 1 $, a limit in which the quarks are ``well localized" and peaked at the boundaries. The coupling constant between the SM quarks and the KK even gauge boson becomes $\sqrt{2}$ times larger than the SM coupling $g_0$.  

In our setup, one may worry about the 5D localized anomaly, which leads to a breakdown of 4D gauge invariance even the zero mode theory, which is the SM, is anomaly free \cite{Scrucca:2001eb}. Because we treat the quarks and leptons separately, the possible 5D anomalies are the $SU(2)_L^2-U(1)_Y$, $U(1)_Y^3$ and $U(1)_Y-$ gravitational anomalies, which do not cancel among the quark or lepton sector alone. For the leptons, the 5D anomalies will live entirely at the boundary $y=\pm L$ (the orbifold fixed plane for the leptons) \cite{ArkaniHamed:2001is}. 
In the interval $y\subset[0, L]$, for a 5D quark field whose left/right-handed component $\Psi_-$/$\Psi_+$ has the $(\alpha_0, \alpha_1)$ boundary condition at $y=0$ and $L$, the 5D anomaly is 
$\partial_C J^C = \mathcal{Q} \Big[ \alpha_0 \delta(y) + \alpha_1 \delta (y-L) \Big] /2$ \cite{Gripaios:2007tk}, where $\mathcal{Q}$ is the corresponding \textsl{consistent anomaly} of a left/right-handed spinor. The 5D anomalies from the KK even states ($\alpha_0 = 1$) will cancel those from the KK odd states ($\alpha_0 = -1$) at the midpoint $y=0$, so the 5D anomalies from the quark sector will live at the boundary $y= L$. Similarly, the 5D anomalies in the interval $y\subset [-L,0]$ will be localized at the boundary $y=-L$. As a consequence, the 5D anomalies from the quark sector will be localized at the same point ($y=\pm L$) as those from the lepton sector, thus our 5D setup is anomaly free.

\noindent {\em  Bounds from EW Data and Flavor Physics--}
 In our setup, the main  one-loop contributions to the electroweak precision parameters from the KK top and bottom are expected to be small as $\delta S, \delta T \sim m_t^2/m_{KK}^2$ \cite{Appelquist:2000nn}. If Higgs is heavy ($m_H \sim 1/R$), contributions to EW precision parameters from the Higgs and its KK modes can be  important \cite{EWPT_MUED}. The additional bounds come from the KK number violation. The oblique parameters will not be affected directly by the KK even gauge bosons  since they will never mix with the $W$ and $Z$ gauge bosons through operators like $(D^\mu H)^{\dag} D_\mu H$ because the Higgs profile is chosen to be flat in UED. The important modification is the four fermion operator which involves the SM quarks by integrating out the KK even gauge bosons. The most stringent bound comes from the resonance search for the Tevatron dijet channels. We consider the ``well localized" case, where couplings between the 2nd KK gluon and the SM quarks is $\sqrt{2} g_s$, the bound for the $g^2$ mass is illustrated in Fig. 2. We can see that the allowed mass region for $g^2$ is $M_{g^2} > 1.3$ TeV, which corresponds to the compactification scale ($R$ is defined as $2 L/ \pi$)
\bea
\label{bounds}
\frac{1}{R} > 510 ~ \rm{GeV} 
\eea
assuming 28\% renormalization group enhancement for the $g^2$ mass \cite{Cheng:2002iz}.

\begin{figure}
\includegraphics[scale=0.43, angle=0]{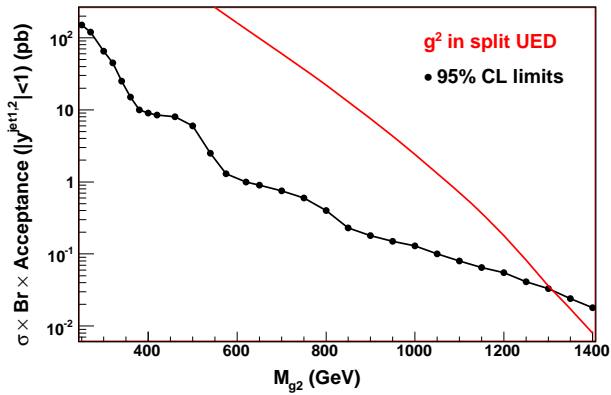}  
\caption{\label{fig2} Comparison of cross section predicted by the ``well localized" scenario of the split-UED and the CDF exclusion bounds on the colored-octet vector boson production cross section as a function of the resonance mass. The $g^2$ production cross section is obtained by rescaling the cross section for the Coloron case in Ref. \cite{TeVdijets} by a factor of 2.} 
\end{figure}

Since we have chosen a flavor universal Yuakwa coupling in Eq. (\ref{eq:Lag}), there is no tree level contributions to the flavor changing neutral currents. The only contributions are from the box diagrams which involves KK gauge bosons, in particular, the KK even weak gauge boson $W_{2n}^{\pm}$, to $K^0-\bar{K^0}$, $B_{s,d}^0-\bar{B}_{s,d}^0$ mixing and $\epsilon_K$. Thanks to the KK parity as well as the orthogonality of the KK decomposition in the flat background, there is no tree level flavor changing neutral current (FCNC) in split-UED, but the KK even weak gauge bosons can contribute to the box diagrams with $(W_{\rm SM}^{\pm},W_{2n}^{\pm})$ exchanges that are suppressed by the mass of $W_{2n}^\pm$:
\begin{eqnarray}
\frac{\Delta M_{s,d}(KK)}{\Delta M_{s,d}(SM)}\simeq \frac{\sum_{n=1}^\infty (2{\cal F}_n) ^2   m_t^2/m_{W_{2n}^2}}{S_0(SM)} < 7.5{\%} \ ,
\end{eqnarray}
where we took $1/R> 510$ GeV as we found above (see Eq.(\ref{bounds})), $S_0(SM)\simeq 2.42$ \cite{Buras} and $m_{W_{2n}}\simeq 2n/R$ as $m_W \ll 2n/R$. In view of non-perturbative uncertanties in $\Delta M_K, \Delta M_{d,s}$ and $\epsilon_K$, it will be very difficult to distinguish the split-UED expectations from the SM ones. In conclusion, the split-UED is safe from the flavor physics bounds.

\noindent {\em Dark Matter--} In our scenario, by making the KK quark mass heavier, the hadronic annihilation cross section could be highly suppressed, as shown in Table \ref{table1}. The relic density of the LKP, which is the first KK photon, will not be affected because the coannihilation of the KK quarks could be safely ignored in the calculation. The precise calculation of the relic density \cite{Kong:2005hn} indicates that the LKP mass has to be from 500 GeV to 700 GeV in the typical mass spectrum \cite{Cheng:2002iz} to get the right relic density, which indeed covers the LKP mass (around 620 GeV)  that is required to fit the peak of electron and positron spectrum \cite{Chen:2009gz} in the ATIC data \cite{ATIC}.

\begin{table}[]
\caption{\label{table1} The branching fraction of the hadronic annihilation of the LKP over that of the charged leptonic annihilation of the LKP, as well as the mass of $q^1$, for different 5D bulk mass $\mu$. Here we fix the LKP mass to be 620 GeV and include the radiative corrections to all the masses according to Ref. \cite{Cheng:2002iz} as an approximation. }
\begin{ruledtabular}
\begin{tabular}[c]{c|c|c|c|c|c|c}
$\mu$ (GeV)& 0   & 200  & 400  & 600 & 800 & 1000\\
   \hline
   $M_{q1}$ (GeV)& 713   & 863  & 1026 & 1198 & 1378 & 1566\\
   \hline
\textrm{BR}(had)/\textrm{BR}(lep) & 45.7\% &  39.3\%  &  28.6\% &  18.3\% & 10.7\% & 6.0\%
\end{tabular}
\end{ruledtabular}
\end{table}

\noindent {\em Collider Physics at the LHC--} Keeping KK parity as in MUED, any KK odd particles should not be singly produced by particle collisions.  
However, the KK even gauge bosons, especially the KK even gluon, will be copiously produced due to the KK number number violation in the valence quark coupling. We expect to discover the resonance in the dijet channels against the large QCD background \cite{Lillie:2007ve} even in the early stages of LHC (assuming an integrated luminosity of $\mathcal{L}=$ 100pb$^{-1}$) \footnote{The production of KK even gauge bosons through loop suppressed couplings in MUED and their discovery through dilepton channel have been discussed in Ref. \cite{Datta:2005zs}.}. To explore this possibility, in Fig. 3, we plot the invariant mass distribution of QCD dijets simulated by Madgraph \cite{Alwall:2007st} (with rough acceptance cuts $ |\eta| < 2.5$ and $p_T>500$ GeV to reduce the SM background). We choose a high minimum $p_T$ cut for the jets so our simulation based on perturbative QCD is reliable. In Table \ref{table2},  we have calculated the decay width and the statistical significance of the signal for the $g^2$ discovery both during the early stage and the entire LHC running time \footnote{The jet resolution is neglected in the simulation because the width of the resonance is much larger.}. We can see that at the early stage, one can discover the second KK gluon $g^2$ just below 4 TeV. For the entire LHC running time, one can clearly discover $g^2$ up to more than 6 TeV \footnote{Here we do not present the statistical significance of the signal for $M_{g^2} > 6$ TeV, because the corresponding $S/B$ is less than 10$\%$, so any deviation from the Gauss distribution of the events will contaminate our estimation.}.

\begin{figure}[]
\includegraphics[scale=0.43, angle=0]{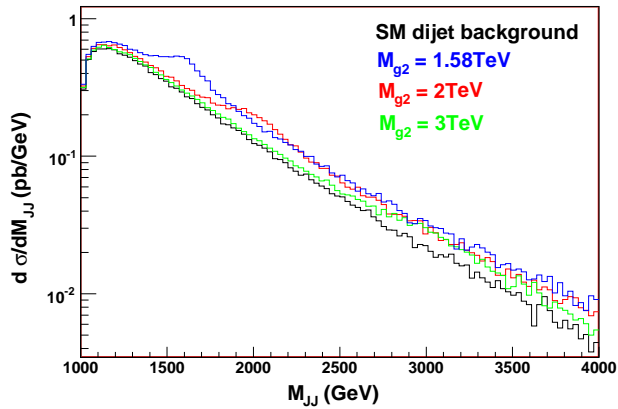}  
\caption{\label{fig3}The invariant mass distribution of the dijet signal. The resonance peaks appear at the
$M_{g_2}$ for 1.58 (for ATIC/PPB-BETS and PAMELA), 2 and 3 TeV. The deviation from the SM background at the high dijet invariant mass comes from the T-channel exchange of the KK even gluon. } 
\end{figure}

\begin{table}[]
\caption{\label{table2} The statistical significance of the signal for the $g^2$ discovery, as well as the $g^2$ decay width with different $g^2$ masses in the ``well localized" scenario. The event number is counted in the mass window $0.85 M_{g^2} < M_{JJ} < 1.2 M_{g^2}$.}
\begin{ruledtabular}
\begin{tabular}[c]{c|c|c|c|c|c|c}
   $M_{g2}$ (TeV)& 1.58   & 2  & 3  & 4 & 5 & 6\\
   \hline
$\Gamma_{g2}$ (GeV) & 270 &  334  &  482 &  627 & 769 & 909\\
\hline
$S/ \sqrt{B}$  (100 pb$^{-1}$) & 66.5 &   38.2   &  11.9 & 4.3 & $-$ & $-$\\
\hline
$S/ \sqrt{B}$ (100 fb$^{-1}$) & 2103 &   1208   &  376 & 137 & 86 & 22
\end{tabular}
\end{ruledtabular}
\end{table}

\noindent {\em Conclusion--}  In the present model, split-UED, the quarks are quasi-localized on the boundaries, but  all of the other fields including leptons, gauge bosons and the Higgs are in the bulk in such a way that KK parity is conserved. We explicitly show how the required properties can be realized in the five dimensional interval and that the set-up is anomaly free. The model is automatically safe from the FCNC problem since all the new interactions are flavor blind. The most stringent bound comes from the resonance search for the Tevatron dijet channels. The LKP is a nice dark matter candidate which can pair annihilate mainly into leptons, thus the model nicely meets the requirements imposed by the recent PAMELA and ATIC/PPB-BETS data. A novel signal at the LHC is the dijet production from a second KK gluon resonance. 

\noindent {\em Acknowledgements--} 
This work was supported by the World Premier International Research Center Initiative 
(WPI initiative) by MEXT, Japan.

\end{document}